\documentclass[aps,prl,groupedaddress,reprint,twocolumn,showpacs,amsmath,amssymb,amsfonts]{revtex4-1}  

\immediate\write18{texcount \jobname.tex -out=\jobname.sum}
\usepackage{graphicx}
\usepackage{dcolumn}
\usepackage{bm}

\usepackage{verbatim}
\usepackage{gensymb}
\usepackage{textcomp}
\usepackage{color}

\newcommand{\CaK}{CaKFe$_4$As$_4$}

\begin{document}
	
	\title{Pressure induced half-collapsed-tetragonal phase in CaKFe$_4$As$_4$ }
	
	\author{Udhara~S.~Kaluarachchi$^{1,2}$, Valentin~Taufour$^{2,*}$, Aashish Sapkota$^{1,2}$,  Vladislav~Borisov$^{3}$,   Tai~Kong$^{1,2,\dagger}$, William~R.~Meier$^{1,2}$, Karunakar~Kothapalli$^{2}$, Benjamin~G.~Ueland$^{2}$, Andreas Kreyssig$^{1,2}$, Roser Valent\'{\i}$^{3}$, Robert~J.~McQueeney$^{1,2}$, Alan~I.~Goldman$^{1,2}$,  Sergey~L.~Bud'ko$^{1,2}$,  Paul~C.~Canfield$^{1,2}$}
	
	\affiliation{$^{1}$Department of Physics and Astronomy, Iowa State University, Ames, Iowa 50011, USA}
	\affiliation{$^{2}$Ames Laboratory, U.S. DOE, Iowa State University, Ames, Iowa 50011, USA}
	\affiliation{$^{3}$Institute of Theoretical Physics, Goethe University Frankfurt am Main, D-60438 Frankfurt am Main, Germany}
	
	\date{\today}
	
	\begin{abstract}
		We report the temperature-pressure phase diagram of CaKFe$_4$As$_4$ established using high pressure electrical resistivity, magnetization and high energy x-ray diffraction measurements up to 6\,GPa.  With increasing pressure, both resistivity and magnetization data show that the bulk superconducting transition of CaKFe$_4$As$_4$ is suppressed and then disappears at $p$\,$\gtrsim$\,4\,GPa. High pressure x-ray data clearly indicate a phase transition to  a collapsed tetragonal phase in CaKFe$_4$As$_4$ under pressure that coincides with the abrupt loss of bulk superconductivity near 4\,GPa. The x-ray data, combined with resistivity data, indicate that the collapsed tetragonal transition line is essentially vertical, occuring at 4.0(5)\,GPa for temperatures below 150\,K.  Band structure calculations also find a sudden transition to a collapsed tetragonal state near 4\,GPa, as As-As bonding takes place across the Ca-layer. Bonding across the K-layer only occurs for $p$\,$\geq$\,12\,GPa. These findings demonstrate a new type of collapsed tetragonal phase in CaKFe$_4$As$_4$: a half-collapsed-tetragonal phase.
		
	\end{abstract}
	

	\maketitle
	Discovery of superconductivity in FeAs-based compounds\,\cite{Kamihara2008, Ren2008, Rotter2008PRL,Takahashi2008Nat} opened a new chapter in high temperature superconductor research. Among them the $AE$Fe$_2$As$_2$ ($AE$=alkaline earth), so called 122 systems, gained attention\,\cite{Canfield2010,Ni2011MRS} due to the ease of growing large, high-quality,  single crystals and the relative simplicity of the structure. At ambient pressure, the parent compounds form in the ThCr$_2$Si$_2$-type structure  and,  in most cases,  undergo structural and magnetic transitions upon lowering of the temperature\,\cite{Canfield2010,Ni2011MRS}. Application of pressure or chemical doping generally suppresses the structural/magnetic transitions and reveals  superconductivity\,\cite{Canfield2010,Ni2011MRS,Torikachvili2008PRL,Torikachvili2008,Alireza2009,Kimber2009Nat,Colombier2009PRB,Sefat2011}. Although chemical substitution is an effective way to alter the lattice parameters and the density of states at the Fermi energy,  it also introduces disorder that can effect the physics in uncontrollable ways. Therefore  physical pressure is one of the cleaner ways to perturb these systems and understand their intertwined and competing phase transitions.
	
	The Ca$^{2+}$ ionic radius in Ca122\,\cite{Canfield2010, Ni2008PRB,Ronning2008JPCM} is the smallest of the $AE$122 family and Ca122 is exceptionally sensitive to  pressure\,\cite{Canfield2010,Torikachvili2008PRL,Kreyssig2008PRB,Yu2009}. Application of less than 0.5\,GPa pressure suppresses the first order structural/magnetic transition and reveals a transition to a non-magnetic, collapsed-tetragonal phase\,\cite{Canfield2010,Torikachvili2008PRL,Kreyssig2008PRB,Yu2009}. In contrast, at ambient pressure K122 shows superconductivity with a relatively low $T_\text{c}$\,\cite{Rotter2008,Sasmal2008PRL,Chen2009} without any structural/magnetic transition. The effect of pressure on K122 has attracted attention due to the unusual "V-shaped"\,\cite{Tafti2013Nat,Terashima2014PRB_b,Taufour2014PRB} pressure dependence of $T_\text{c}$ with a minimum in $T_\text{c}$ near 1.8\,GPa.  At higher pressures, a collapsed-tetragonal phase was observed for  $p$\,$\geq$\,16\,GPa at room temperature\,\cite{Nakajima2015,Ying2015arXiV,Wang2016PRBb,Guterding2015PRB}. 
	
	Recently a new  superconductor with a related, structure, Ca$A$Fe$_4$As$_4$ ($A$=K,Rb,Cs), was discovered by Iyo ${et~al.}$\,\cite{Iyo2016JACS}. Unlike a (Ca$_{1-x}$K$_x$)122\,\cite{Wang2013b} solid solution, which has the I4/$mmm$ body-centered-tetragonal space group, CaKFe$_4$As$_4$\,\cite{Iyo2016JACS,Meier2016PRB} has separate, unique crystallographic sites for the alkaline and alkaline-earth metals, and possesses in the primitive tetragonal P4/$mmm$ space group. At ambient pressure, CaKFe$_4$As$_4$ shows bulk superconductivity below $T_\text{c}$$\sim$\,35\,K in zero applied field or below 700\,kOe\,$\leq$\,$H_\textrm{c2}$\,$\leq$\,900\,kOe at low temperatures\,\cite{Meier2016PRB}. Initial pressure work up to $\sim$\,4\,GPa\,\cite{,Meier2016PRB} has shown that $T_\text{c}$ is suppressed to 28.5\,K by 3.9\,GPa. To further determine and understand the $p$-$T$ phase diagram of this system, higher-pressure studies are needed.
	
	In this Letter we present resistivity, magnetization and structural measurements of CaKFe$_4$As$_4$ under pressures up to 6\,GPa and find a half-collapsed-tetragonal (hcT) phase stabilized for $p$\,$\gtrsim$\,4\,GPa. Above this pressure there is an $\approx$\,2.6\,$\%$ decrease in the $c$-axis lattice parameter and an $\approx$\,0.4\,$\%$ increase in the $a$-lattice parameter. This half-collapsed tetragonal phase transition coincides with the abrupt loss of bulk superconductivity in the magnetization and the resistance measurements. Band structure calculations show that, for $p$\,$\gtrsim$\,4\,GPa, As-As bonding takes place across the Ca-layer  with much higher pressures needed to cause  such bonding to span the larger K-layer. Taken together CaKFe$_4$As$_4$ undergoes a simultaneous structural collapse and loss of superconductivity with As-As bond formation across the Ca- but not K- layers. 
	
	Ca$K$Fe$_4$As$_4$ single crystals were grown using high-temperature solution growth technique described in Ref\,\onlinecite{Meier2016PRB}. Temperature dependent magnetization and resistance were carried out using a Quantum Design(QD) Magnetic Property Measurement System(MPMS) and a QD Physical Property Measurement System(PPMS) respectively. The ac resistivity was measured by the standard four-probe method with the current in the $ab$ plane. Four Pt wires, with  diameters of 25\,$\mu$m, were soldered to the sample using a Sn:Pb-60:40 alloy. For the resistivity measurements, pressure was applied at a room temperature using a modified Bridgman cell\,\cite{Colombier2007} with a 1:1 mixture of $n$-pentane:iso-pentane  as a pressure medium\,\cite{Piermarini1973}. The pressure was determined at low temperature by monitoring the superconducting transition temperature of Pb\,\cite{Bireckoven1988,Eiling1981}. High pressure magnetization measurements were carried out using a moissanite anvil cell\,\cite{Alireza2007} with Daphne 7474 as a pressure medium\,\cite{Murata2008}. The pressure was applied at a room temperature and  the ruby fluorescence technique\,\cite{Piermarini1975} at 77\,K was used to determine the pressure. See Supplemental Material\,\cite{Supplement} for further details.
	
	High-energy x-ray diffraction measurements were performed on a six-circle diffractometer at end station 6-ID-D at the Advanced Photon Source, using an x-ray energy of $E$\,=\,100.33\,keV and a beam size of 100$\times$100\,$\mu$m$^2$. Single-crystal samples were loaded into a double-membrane-driven\,\cite{Sinogeikin2015RSI} copper-beryllium diamond-anvil cell (DAC). Helium was used as the pressure-transmitting medium and loaded to a pressure of 0.7\,GPa at 300\,K. Ruby spheres and silver foil were also mounted in the DAC for pressure determination. The DAC was attached to the cold finger of a He closed-cycle refrigerator. Diffraction patterns were recorded using a MAR345 image plate detector positioned at 1.494\,m from the sample position. The distance was determined from measurement of powder patterns of a CeO$_2$ standard from the National Institute of Standards and Technology. The detector was operated with a pixel size of 100$\times$100\,$\mu$m$^2$, and patterns were recorded while rocking the sample through two independent angles up to $\pm$\,3.6$\degree$ about the axes perpendicular to the incident beam. The measurement was performed in the (H, K, H) scattering plane\,\cite{Supplement}.$~$$~$
	
	In order to investigate the electronic and structural properties of CaKFe$_4$As$_4$, we performed density functional theory (DFT) relativistic calculations using the Vienna Ab initio Simulation Package (VASP)\,\cite{Kresse1993PRB,Kresse1996PRB,Kresse1996} with the projector-augmented wave (PAW) basis\,\cite{Blochl1994PRB,Kresse1999PRB} in the generalized-gradient approximation (GGA). In order to take into account, in the first approximation, the paramagnetic fluctuations that preserve the tetragonal symmetry of the lattice ($a = b$), we consider a model with a ``frozen'', twisted, long-range magnetic order which is one of the lowest-energy configurations with this symmetry. Pressure-dependent structures were obtained by fixing the components of the stress tensor to the given value (equal to the external pressure) and fully relaxing the lattice parameters and the internal atomic positions with the conjugate-gradient method\,\cite{Tomic2012PRB,Tomic2013PRB,Dhaka2014PRB}. The integration over the irreducible Brillouin zone was realized on the $\Gamma$-centered $(10\times 10\times 10)$ k-mesh\,\cite{Supplement}.

	\begin{figure}[b!]
		\begin{center}
			\includegraphics[width=85mm]{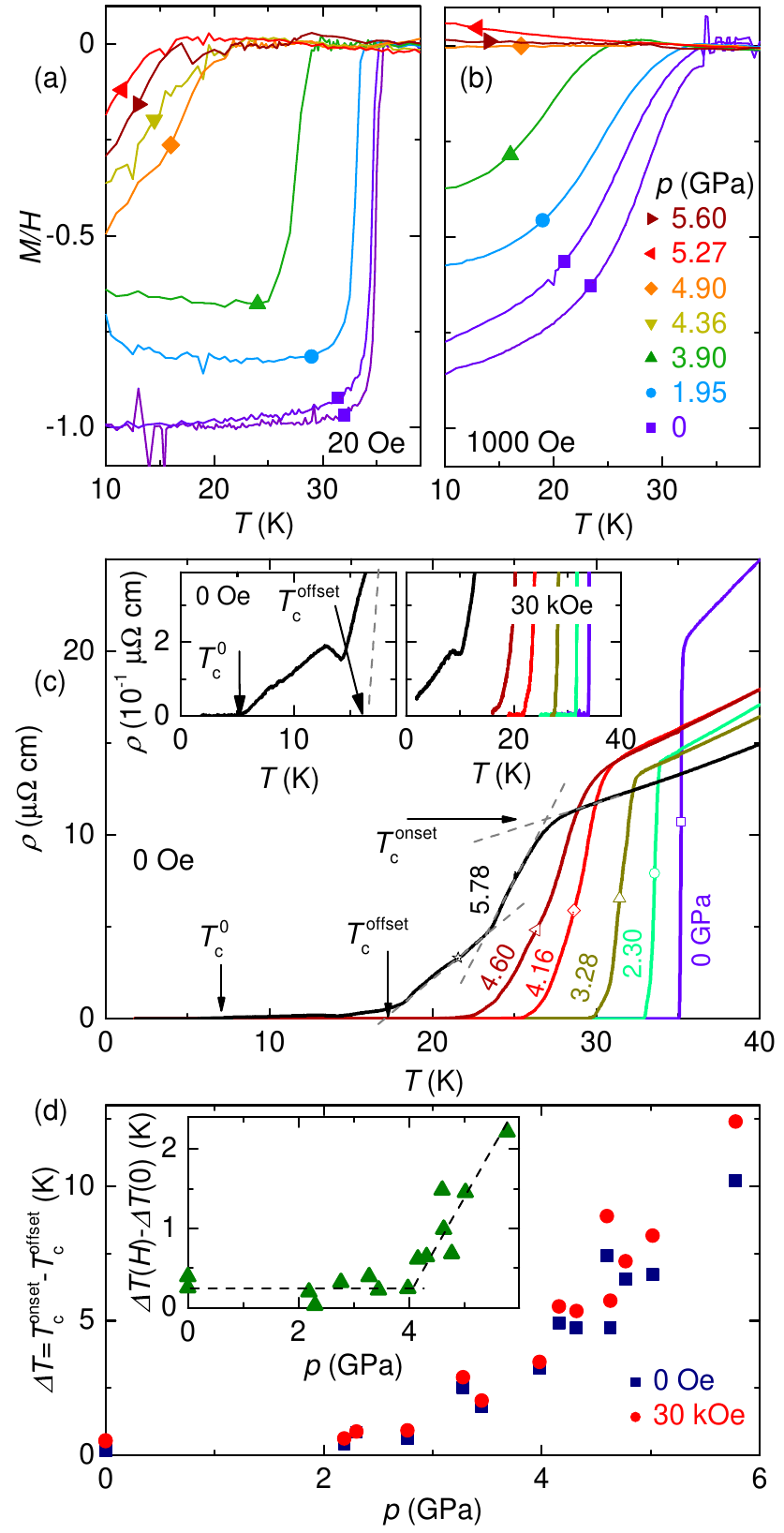}
		\end{center}
		\caption{(Color online)  The zero-field-cooled(ZFC) $M$($T$) data for (a) 20\,Oe and (b) 1000\,Oe applied field. (c) The temperature dependent resistivity at representative pressures. Arrows for the 5.78\,GPa curve indicate $T_\text{c}^\text{onset}$, $T_\text{c}^\text{offset}$ and $T_\text{c}^\text{0}$ criteria used in this paper. The left and right insets show the temperature dependent resistivity under pressure for 0\,Oe and 30\,kOe applied fields respectively. (d) Superconducting transition widths, $\Delta T$\,=\,$T_\text{c}^\text{onset}-T_\text{c}^\text{offset}$, for 0 and 30\,kOe. The inset shows the difference of transition width,  $\Delta T\text{(H)}-\Delta T\text{(0)}$, where  $\Delta T\text{(H)}$ is the transition width  at 30\,kOe}\label{LowT_Phase}
	\end{figure}
	
	Figures\,\ref{LowT_Phase}\,(a)\,-\,(b) show the zero-field-cooled (ZFC) magnetization data, $M(T)$, on single crystaline samples of CaKFe$_4$As$_4$  under pressure below 40\,K. These data have been normalized to the ZFC data of a larger sample at ambient pressure\,\cite{Meier2016PRB}. The low field $M(T)$ data in Fig.\,\ref{LowT_Phase}\,(a) show a clear diamagnetic signal below $T_\textbf{c}$. $T_\textbf{c}$ monotonically decreases with the pressure up to 3.9\,GPa. Upon further increase of pressure, there is a broader and reduced diamagnetic anomaly  below 20\,K. The appearance of a small diamagnetic signal above 3.9\,GPa can be related to non-bulk/filamentary superconductivity and can  often be easily eliminated by the application of a small (1000\,Oe\,$\ll$\,$H_\textrm{c2}$)\,\cite{Meier2016PRB} magnetic field (Fig.\,\ref{LowT_Phase}\,(b)) where the observation of a clear diamagnetic signal can be found only up to 3.9\,GPa. This suggests non-bulk superconductivity for pressures above 3.9\,GPa.

	A similar decrease in the superconducting transition temperature is observed in the resistivity measurements under pressure (Fig.\,\ref{LowT_Phase}\,(c)). The sharp superconducting transition found at low pressures becomes broadened at higher pressures. The superconducting transition temperature is parameterized by three different criteria, $T_\text{c}^\text{onset}$, $T_\text{c}^\text{offset}$ and $T_\text{c}^\text{0}$  as shown by arrows in Fig.\,\ref{LowT_Phase}\,(c) and its insets. In contrast to the loss of superconducting signature in $M(T)$, $\rho(T)$ shows zero resistivity up to 5.78\,GPa. This is due to the fact that resistivity measurements are known to be more susceptible to filamentary superconductivity, thus, application of a magnetic field helps to separate it from bulk superconductivity. The left and right insets of Fig\,\ref{LowT_Phase}\,(c) show the temperature dependent resistivity under pressure for 0\,Oe and 30\,kOe fields. The clear observation of non-zero resistivity at 5.78\,GPa at 30\,kOe again confirms the non-bulk nature of the higher pressure superconductivity. 
	
	In order to determine the transition from bulk to non-bulk superconductivity in the electrical resistivity data we focus on the transition widths. The superconducting transition width, $\Delta T$\,=\,$T_\text{c}^\text{onset}-T_\text{c}^\text{offset}$, is broadened with the increase of pressure as illustrated by Fig.\,\ref{LowT_Phase}\,(d), but this can be, in part, due to pressure inhomogeneities in the pressure medium, rather than represent any sort of phase transition. In order to search for a possible transition we compare zero field and moderate, but finite field ($H$\,=\,30\,kOe\,$\ll$\,$H_\textrm{c2}$) resistive transition width data. Any broadening due to inhomogeneous pressure should be equally present in the $H$\,=\,0\,Oe and $H$\,=\,30\,kOe data. To determine the field contribution of the transition width, we calculated the difference of the transition widths, $\Delta T\text{(H)}-\Delta T\text{(0)}$ (inset of  Fig.\,\ref{LowT_Phase}\,(d)). $\Delta T\text{(H)}-\Delta T\text{(0)}$ remains constant until near 4\,GPa and then starts to increase significantly for pressures above that. This critical pressure is consistent with the pressure at which the magnetization data shows a loss of  bulk superconductivity. Although broadening of the superconducting transition in field can be also due to the thermal fluctuation of the vortex system, the observation, in magnetization data, of a loss of diamagnetic signal at 1000\,Oe eliminates this possibility.
	
	The data in Fig.\,\ref{LowT_Phase}  can be summarized in a pressure dependence of $T_\textbf{c}$ phase diagram as shown in Fig.\,\ref{Phase_diag}. The solid and open symbols represent the bulk and non-bulk superconductivity. Bulk superconductivity persists up to a critical pressure of $\sim$\,4\,GPa. The faint signatures of  superconductivity appearing above 4\,GPa in this phase diagram are due to strain-induced, filamentary superconductivity. The sudden and discontinuous disappearance of superconductivity for $p$\,$\geq$\,4\,GPa is very suggestive of what is found for Ca(Fe$_{1-x}$Co$_x$)$_2$As$_2$ both as a function of $x$\,\cite{Ran2012PRB} and as a function of $p$\,\cite{Gati2012PRB}, i.e. superconductivity suddenly and discontinuously disappears when a collapsed tetragonal, non-magnetic phase is stabilized
	
	\begin{figure}[tb!]
		\centering
		\begin{center}
			\includegraphics[width=85mm]{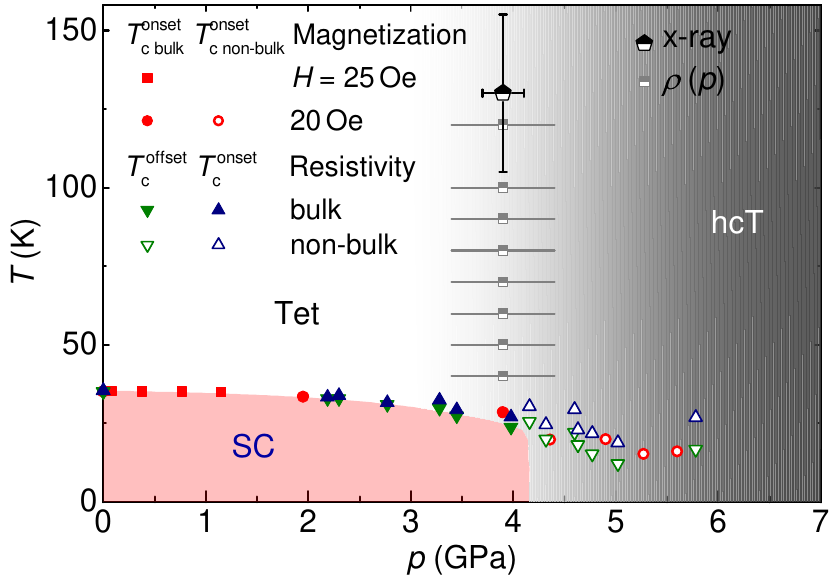}
		\end{center}
		\caption{(Color online) Temperature-pressure phase diagram of CaKFe$_4$As$_4$, constructed from magnetization, resistivity and x-ray diffraction measurements under pressure.  Solid and opened symbols represent $T_\text{c}$ for bulk and filamentary superconductivity  respectively. Red color square symbols are from Ref\,\onlinecite{Meier2016PRB}. The green and blue color data represent $T_\text{c}^\text{offset}$ and $T_\text{c}^\text{onset}$ respectively. The light-red color shaded area represent the bulk superconductivity region. Half filled pentagon and squares represent the data from high pressure x-ray diffraction and the resistivity anomaly as seen in Fig.\,\ref{latticepara_p}\,(a) and Fig.\,\ref{Rho_p_isotherm} }\label{Phase_diag}
	\end{figure}

	\begin{figure}[b!]
		\begin{center}
			\includegraphics[width=85mm]{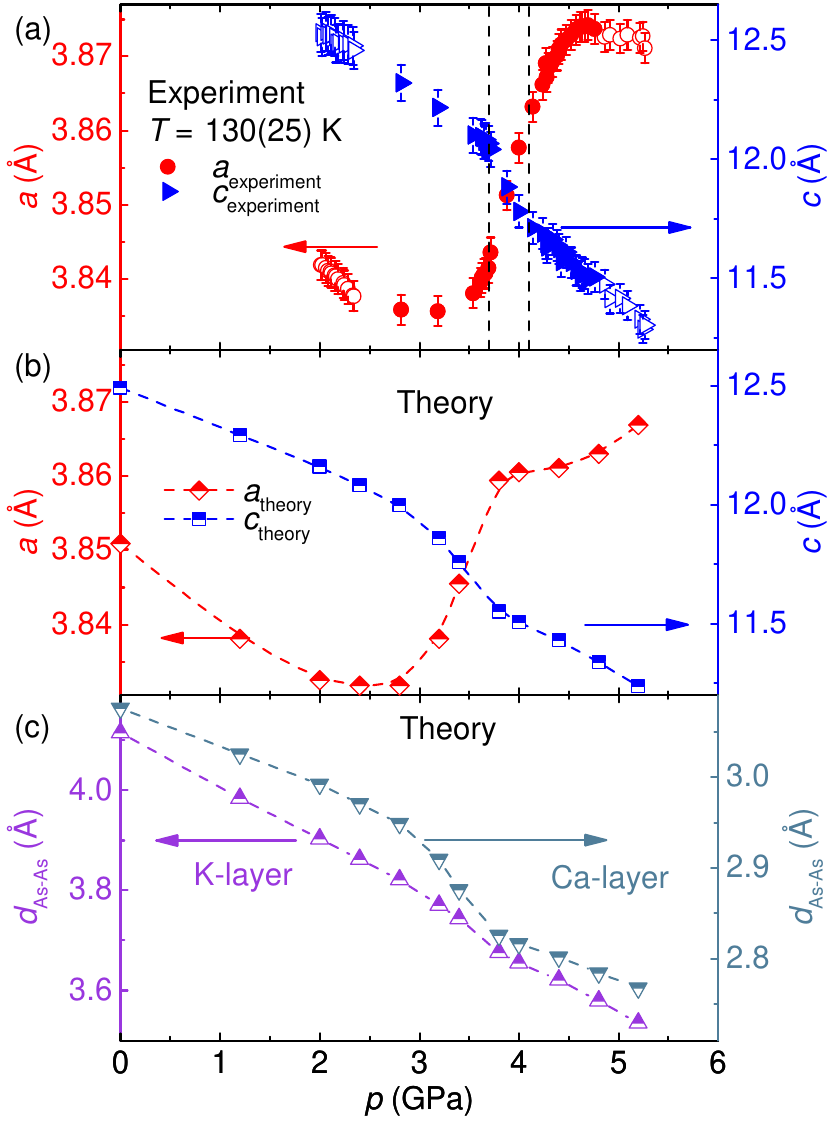}
		\end{center}
		\caption{(Color online)   (a) Pressure dependent $a$  (left axis) and $c$ (right axis) lattice parameters obtained from the high pressure x-ray diffraction study. The solid symbols represent the data measured within the temperature range of 130(25)\,K and open symbols are for higher or lower temperature than this range. The lattice parameters are obtained after fitting the one-dimensional (0,2,0) and (1,0,1) peaks. The one-dimensional peak is obtained from two-dimensional pattern after selecting the region of interest around the peak and integrating azimuthally. The dashed-verticle lines markes the bound of the transition and represent as the error bar in Fig.\,\ref{Phase_diag}. (b) Theoretical $a$  (left axis) and $c$ ( right axis) lattice parameters represented with the same scale as (a). (c) Change of the As-As distances across the K (left axis) and Ca (right axis) cationic layers under the transition to the collapsed tetragonal phase. Dashed lines are guides for the eye.} \label{latticepara_p}
	\end{figure}

	To further pursue these findings, we examined the pressure dependence of the lattice parameters from the diffraction study for temperatures near 130\,K (Fig.\,\ref{latticepara_p}\,(a)). We observed an abrupt enhancement of the $a$-lattice parameter simultaneous to a significant reduction of the $c$-lattice parameter near 4\,GPa without any crystallographic symmetry change. This indicates a pressure induced phase transition from a tetragonal to a collapsed tetragonal phase, similar to the observation in other 122 systems\,\cite{Kreyssig2008PRB,Goldman2009,Nakajima2015,Ying2015arXiV,Uhoya2010PRB,Mittal2011PRB,YU2014SR,Uhoya2010JPCM,Uhoya2011JPCMb,Uhoya2011JPCM}. The change of the  $a$-lattice parameter is about $\approx$\,0.016\AA\,($0.4\%$) ~while the $c$-lattice parameter decreases by $\approx$\,0.31\AA\,($2.6\%$).

	Band structure calculations manifest a very similar expansion in $a$ of $\approx$\,0.015\AA\,($0.4\%$) while  $c$ drops by $\approx$\,0.17\AA\,($1.5\%$) at the critical pressure of 4\,GPa as shown in Fig.\,\ref{latticepara_p}\,(b).  As can be seen in Fig.\,\ref{Phase_diag}, this collapsed-tetragonal transition point is positioned directly above the disappearance of bulk superconductivity at lower temperatures.
	
	Analysis of the orbital-resolved band structures at different pressures near this transition confirms its collapsed nature\,\cite{Supplement}. The As antibonding molecular orbitals shift above the Fermi energy upon crossing 4~GPa (Figs.\,\ref{f:collapse_near_Ca_and_K} in the Supplemental Material\,\cite{Supplement}). This is in accordance with a jump-like reduction of the As-As bond length across the Ca layer by $\sim 0.05\:\mathrm{\AA}$ (Fig.\,\ref{latticepara_p}(c)) and an increase in the As-$4p_z$ electron density between the two As ions (Fig.\,\ref{f:collapse_near_Ca_and_K} in the Supplemental Material\,\cite{Supplement}). A second collapse  associated with As-As bonding across the K-layer is found for $p$\,$\sim$\,12\,GPa\,\cite{Supplement}. The transition at $\sim$\,4\,GPa then is associated with a half-collapsed-tetragonal (hcT) phase.

	\begin{figure}[b!]
		\centering
		\begin{center}
			\includegraphics[width=85mm]{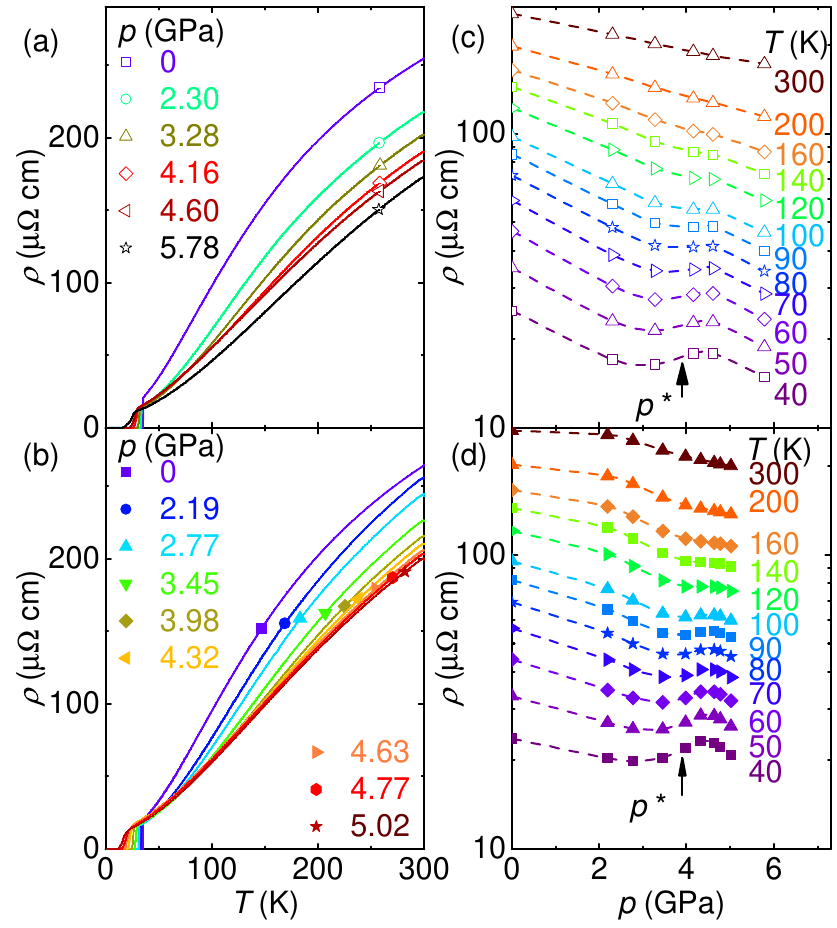}
		\end{center}
		\caption{(Color online) (a)\,-\,(b) Evolution of the temperature dependent resistivity of CaKFe$_4$As$_4$ with pressure for two samples. (c)\,-\,(d) Pressure dependence of resistivity at fixed temperature for same two samples. The arrow indicates the observed resistivity anomaly at $p^*$. }\label{Rho_p_isotherm}
	\end{figure}
	
	Whereas our scattering data point is for $T$\,=\,130(25)\,K, the band structure calculations are formally for $T$\,=\,0\,K and our $T_\textrm{c}$ data is for $T$\,$\sim$\,30\,K. This begs the question of whether a half-collapsed-tetragonal phase line can be detected for $p\sim$\,4\,GPa below 150\,K.  Figures\,\ref{Rho_p_isotherm}\,(a)\,-\,(b) show temperature dependent  resistivity, under various pressures, data sets measured from 300\,K to 1.8\,K. At ambient pressure  CaKFe$_4$As$_4$ shows metallic behavior and becomes superconducting below 35\,K. With increase of pressure, the resistivity at room temperature monotonically decrease up to 5.6\,GPa. However, the resistivity data at lower temperature, in the  normal state ($\geq$\,40\,K), show a non-monotonic pressure evolution. No clear anomaly is visible in the temperature dependent  resistivity, suggesting that any phase transition line will be near vertical. To see this transition line more clearly, we plotted resistivity vs pressure at fixed temperatures (Figs.\,\ref{Rho_p_isotherm}\,(c)\,-\,(d)). At higher temperatures,  the resistivity monotonically decreases with pressure. However, below 160\,K we see a jump in resistivity developing near $p^*$\,$\sim$\,4\,GPa. This can be related to the collapsed tetragonal phase transition as seen in high pressure x-ray measurement. Similar features in the resistivity have been observed in K122\,\cite{Wang2016PRBb} with no clear features in $\rho(T)$ data, but V-shape minimum in $\rho$($p$).

	The composite $p-T$ phase diagram of CaKFe$_4$As$_4$ is shown in the Fig.\,\ref{Phase_diag}. The $p^*$($T$) data inferred from the Figs.\,\ref{Rho_p_isotherm}\,(c)\,-\,(d) ($p^*$ being mid point of rise in $\rho$($p$) data and error bars being width of rise) clearly links the higher-temperature  x-ray data to the lower temperature loss of superconductivity and the $T$\,=\,0 band structure results. Figure\,\ref{Phase_diag} shows that the transition of Tet\,$\rightarrow$\,hcT is nearly vertical.


	To summarize, the bulk superconducting transition temperature of tetragonal CaKFe$_4$As$_4$ is gradually suppressed under pressure up to $\sim$\,4\,GPa and then discontinuously disappears. High pressure x-ray measurements combined with band structure calculations identify the phase transition from a tetragonal to a half-collapsed tetragonal phase near 4\,GPa. From the diffraction study at 130\,K, we find an abrupt enhancement in a-lattice parameter ($0.4\%$) and a reduction in c-lattice parameter ($2.6\%$) near 4\,GPa. Band structure calculations show  that the first collapse occurs on across the Ca-layer near 4\,GPa followed by a second collapse, on across the K-layer, at higher pressures. Both transitions are accompanied by the typical enhancement of the corresponding As-As bonding. Although only half of the As-As bonding is collapsing at 4\,GPa, when the superconducting state is fully suppressed.

	We would like to acknowledge discussions with Peter Hirschfeld, Rafael Fernandes, Milan Tomi\'c, Wageesha Jayasekara and thank Daniel Guterding for providing a code\,\cite{pytricubic} to generate the As-$4p_z$ electron density maps from the Wannier function analysis and D.\,S.\,Robinson for support during the x-ray experiments. Experimental work was supported by the U.S. Department of Energy (DOE), Office of Science, Basic Energy Sciences, Materials Science and Engineering Division and was performed at the Ames Laboratory, which is operated for the U.S. DOE by Iowa State University under contract No. DE-AC02-07CH11358. W. M. were supported by the Gordon and Betty Moore Foundations EPiQS Initiative through Grant GBMF4411. V. T. is supported by Ames Laboratory’s laboratory-directed research and development (LDRD) funding for magnetization measurements under pressure. This research used resources of the Advanced Photon Source, a U.S. Department of Energy (DOE) Office of Science User Facility operated for the DOE Office of Science by Argonne National Laboratory under Contract No. DE-AC02-06CH11357.  The theoretical work  was financially supported by the German Research Foundation (Deutsche Forschungsgemeinschaft) through grant SFB/TR49. The computer time was allotted by the center for supercomputing (CSC) in Frankfurt.
	\newline
	%

	
	\noindent$*$ Current affiliation: Department of Physics, University of California, Davis, California 95616, USA.
	\newline
	$\dagger$ Current affiliation: Department of Chemistry, Princeton University, Princeton, NJ 08544, USA.


\maketitle

\clearpage
\setcounter{figure}{0}
\renewcommand{\thefigure}{S\arabic{figure}}

\section{Supplementary Material}
\section{Experimental Methods}

Single crystals were synthesized by the high-temperature solution growth technique described in Ref\,\onlinecite{Meier2016PRB}. The experiments under pressure were performed on two samples (samples 1 and 2) in magnetization cells and on another two samples (samples 3 and 4) in resistivity cells. Temperature dependent magnetization and resistivity were carried out using a Quantum Design(QD) Magnetic Property Measurement System(MPMS) and a QD Physical Property Measurement System(PPMS) respectively.

The ac resistivity ($f$\,=\,17\,Hz) was measured by the standard four-probe method with the current ($I$\,=\,1\,mA) in the $ab$ plane. The samples used for the resistivity measurement typically had dimension of 500$\times$100$\times$20\,$\mu$m$^3$ ($l$$\times$$w$$\times$$t$). Four Pt wires, with  diameters of 25\,$\mu$m, were soldered to the sample using a Sn:Pb-60:40 alloy. A modified Bridgman cell\,\cite{Colombier2007} was used to generate pressure for the resistivity measurements. A 1:1 mixture of $n$-pentane:iso-pentane was used as a pressure medium\,\cite{Piermarini1973}; the solidification of this medium occurs around $\sim$6-7\,GPa at room temperature\,\cite{Piermarini1973,Klotz2009,Tateiwa2010,Kim2011PRB,Torikachvili2015RSI}. The pressure was determined at low temperature by monitoring the superconducting transition temperature of Pb\,\cite{Bireckoven1988,Eiling1981}. 

High pressure magnetization measurements were carried out using a moissanite anvil cell\,\cite{Alireza2007}. Samples with dimensions of 100$\times$100$\times$20\,$\mu$m$^3$ were loaded into the cell and Daphne 7474 was used as a pressure medium\,\cite{Murata2008}; the solidification of this medium occurs around 3.7\,GPa at room temperature\,\cite{Murata2008}. The pressure was determined by the ruby fluorescence technique\,\cite{Piermarini1975} at 77\,K. We first collected the raw SQUID voltage of a hand-tight pressure cell with the gasket, pressure medium and ruby chip. This background was measured between 10 to 40\,K in steps of 0.5\,K at 20\,Oe and 1000\,Oe with a demagnetization prior each change of magnetic field. A similar set of measurements were performed with a sample pressurized in the cell. For all sets of measurements, 3 scans (each 32 points over 4\,cm) were recorded and the average was calculated. A point-by-point subtraction of the raw SQUID voltage background was done with the help of a computer program that allows for the rejection of non-reproducible scans. A thermalization time of one to two minutes was necessary to obtain reproducible scans due to the rather large mass of the pressure cell. A fit of the resulting signal was used to extract the magnetization\,\cite{QDMPMS}. To correct for a small shift in position
between the various sets of measurements (various pressures and background), a shift smaller than 1\,mm was used. The shift was determined so as to give the best fit of the sample signal at the lowest temperature (for the largest superconducting signal from the sample). The position of the sample obtained from this scan was then kept as a fixed parameter for the measurements at various temperature and magnetic field values.

High-energy x-ray diffraction measurements were performed on a six-circle diffractometer at end station 6-ID-D at the Advanced Photon Source, using an x-ray energy of $E$\,=\,100.33\,keV and a beam size of 100$\times$100\,$\mu$m$^2$. Single-crystal samples with dimensions of 40$\times$40$\times$25\,$\mu$m$^3$ were loaded into a double-membrane-driven\,\cite{Sinogeikin2015RSI} copper-beryllium diamond-anvil cell (DAC). Helium was used as the pressure-transmitting medium and loaded to a pressure of 0.7\,GPa at 300\,K. Ruby spheres and silver foil were also mounted in the DAC for pressure determination. The DAC was attached to the cold finger of a He closed-cycle refrigerator, and measurements for various values of $p$ for $T$\,=\,130(25)\,K. At this temperature the He pressure medium solidifies near 3.5\,GPa\,\cite{Maury1979,Datchi2000PRB}. As a result there can be some broadening of structural phase transitions for higher pressures than this. Diffraction patterns were recorded using a MAR345 image plate detector positioned at 1.494\,m from the sample position. The distance was determined from measurement of powder patterns of a CeO$_2$ standard from the National Institute of Standards and Technology. The detector was operated with a pixel size of 100$\times$100\,$\mu$m$^2$, and patterns were recorded while rocking the sample through two independent angles up to $\pm$\,3.6$\degree$ about the axes perpendicular to the incident beam. The measurement was performed in the (H, K, H) scattering plane. It should be noted that, although there was finite temperature drift during measurements under pressure (e.g. $T$\,=\,130(25)\,K), the observed behavior in lattice parameters (Fig.\,\ref{latticepara_p}\,(a)) is solely due to pressure rather than any possible temperature variation given that the change in both lattice parameters in the above mentioned temperature range at ambient pressure is almost negligible\,\cite{Meier2016PRB}(temperature dependence of $a$ and $c$ lattice parameters at 130\,K are 4$\times$10$^{-5}$\,\AA\,K$^{-1}$ and 6.8$\times$10$^{-4}$\,\AA\,K$^{-1}$ respectively). Moreover, the $a$-lattice parameter should decrease with the decrease in temperature.

\section{Computational details and results}

In order to investigate the electronic and structural properties of CaKFe$_4$As$_4$, we performed density functional theory (DFT) relativistic calculations using the Vienna Ab initio Simulation Package (VASP)\,\cite{Kresse1993PRB,Kresse1996PRB,Kresse1996} with the projector-augmented wave (PAW) basis\,\cite{Blochl1994PRB,Kresse1999PRB} in the generalized-gradient approximation (GGA). In order to take into account, in the first approximation, the paramagnetic fluctuations that preserve the tetragonal symmetry of the lattice ($a = b$), we consider a model with a ``frozen'', twisted, long-range magnetic order which is one of the lowest-energy configurations with this symmetry (see discussion below). Pressure-dependent structures were obtained by fixing the components of the stress tensor to the given value (equal to the external pressure) and fully relaxing the lattice parameters and the internal atomic positions with the conjugate-gradient method\,\cite{Tomic2012PRB,Tomic2013PRB,Dhaka2014PRB}. The integration over the irreducible Brillouin zone was realized on the $\Gamma$-centered $(10\times 10\times 10)$ k-mesh.

\subsection{A) Magnetic configuration}

\begin{figure}[t!]
	\begin{center}
		\includegraphics[width=85mm]{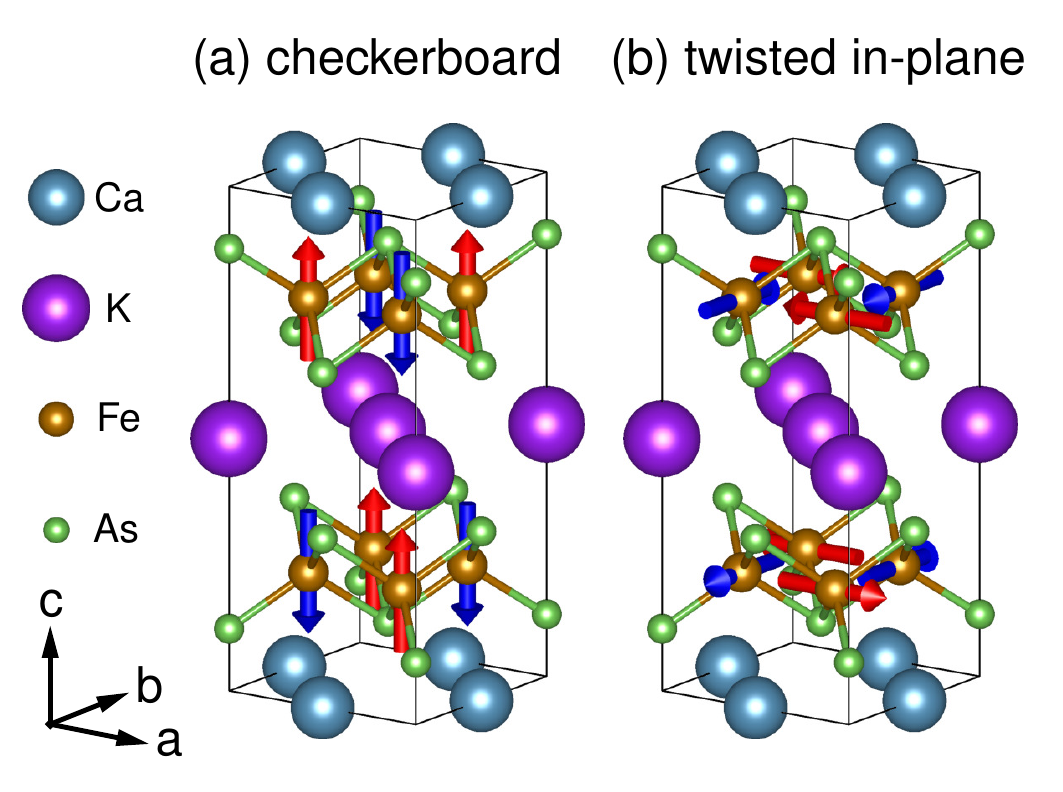}
	\end{center}
	\caption{(a) The collinear checkerboard and (b) the twisted in-plane configuration of the Fe magnetic moments in \CaK{}.}
	\label{f:supercells}
\end{figure}

Since the experimentally observed ambient pressure tetragonal \CaK{} manifests no signature of a structural or magnetic transition but does manifest a clear superconducting transition near 35\,K, we consider first the influence of pressure on the tetragonal non-magnetic phase. In addition, given that paramagnetic fluctuations generally play a role for the structural features of iron pnictides,  we also simulated such fluctuations, to a first approximation, by imposing a ``frozen'' long-range magnetic order which, on the one hand, assists the structural collapse and, on the other hand, preserves the tetragonal symmetry of the crystal lattice. In this respect, the usual stripe-ordered phase is not suitable for these simulations, as it breaks the tetragonal symmetry. Recent works\,\cite{Fernandes2016PRB,Scherer2016PRB} propose a few candidates for the magnetic order in iron pnictides, including spin vortex and charge-spin density wave orders, which have the $C_4$ symmetry opposed to the conventional $C_2$ order often found in the \textit{A}Fe$_2$As$_2$ compounds. In addition to these, we considered a few other magnetic configurations, including the checkerboard arrangement (Fig.~\ref{f:supercells}\,(a)), and found that one of the lowest-energy states is a twisted phase with Fe moments lying in-plane within each FeAs layer (Fig.~\ref{f:supercells}\,(b)). While the Fe moments in each FeAs layer are twisted around the \textit{c} crystallographic axis, the nearest Fe neighbors from the adjacent layer are oriented in the opposite direction. Here, the local symmetry of each Fe site is orthorhomic, since the spins are oriented either along the \textit{a} or along the \textit{b} axis, while the overall unit cell symmetry is tetragonal. Marginal deviations ($<2\cdot 10^{-5}$) from the 90\textdegree{}-tetragonal lattice are observed in our calculations most likely due to the spin-lattice coupling. We note that, if the structure is not restricted to the tetragonal symmetry class, the stripe phase lowers the energy. Nevertheless, in order to agree with the experimentally determined crystal symmetry of \CaK{}\,\cite{Meier2016PRB}, we study in more details the proposed twisted phase (Fig.~\ref{f:supercells}\,(b)) as an approximate model of paramagnetic fluctuations.

\begin{figure*}[tp!]
	\begin{center}
		\includegraphics[width=160mm]{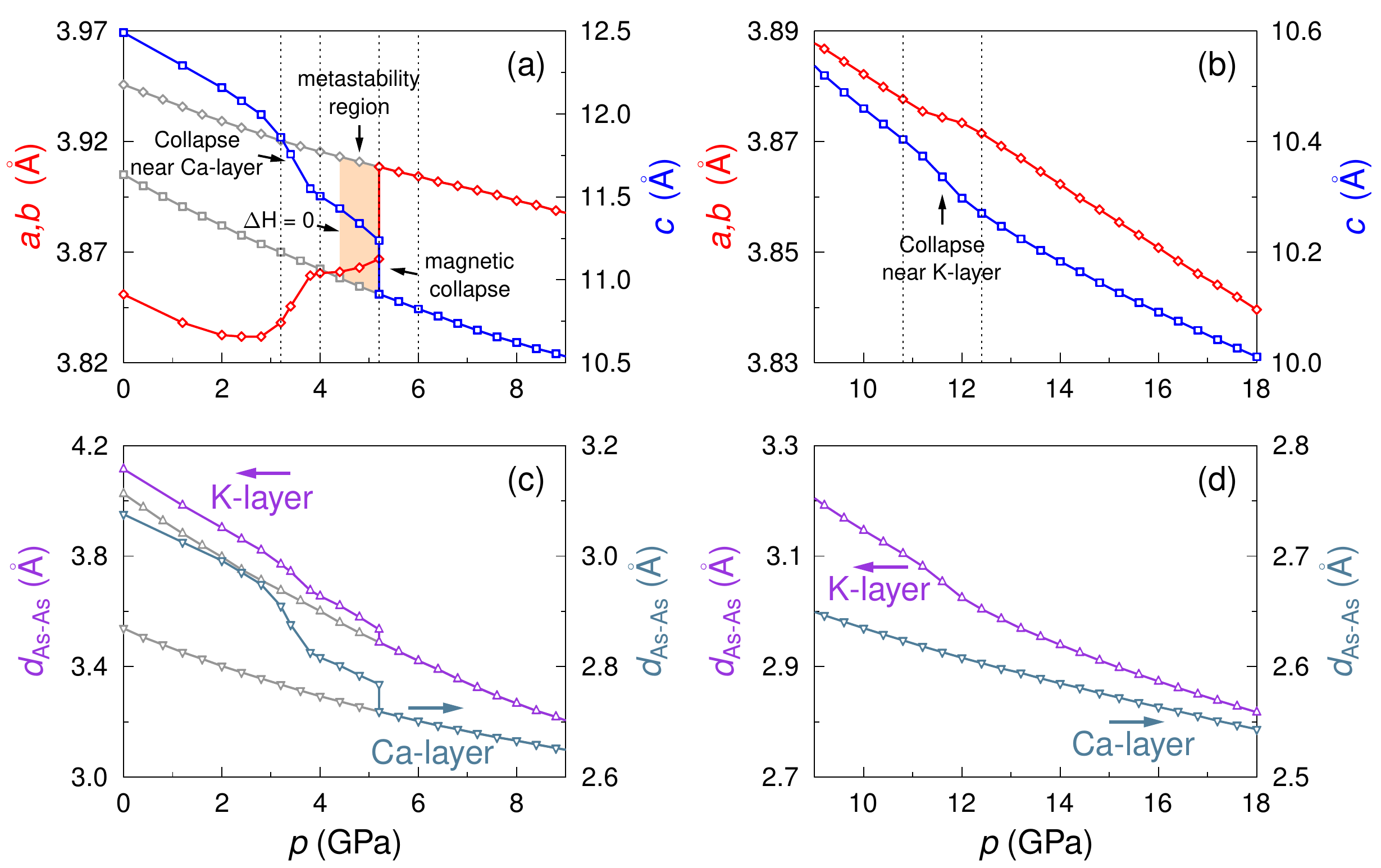}
	\end{center}
	\caption{(a,b) Pressure-dependent lattice parameters and (c,d) the two types of As-As distances in \CaK{} are presented for the twisted magnetic and the non-magnetic (a,c) phases. In plots (a) and (b), transitions around 4~GPa, 5.2~GPa, and 12~GPa are marked with arrows and text labels. Magnetic metastability region is indicated by the light-shaded region between 4.5~GPa and 5.2~GPa. Vertical dashed lines show the values of pressure closest to these transitions at which the electronic properties were analyzed (Fig.~\ref{f:collapse_near_Ca_and_K}). In the background of plots (a) and (c), the low-pressure behavior of the tetragonal non-magnetic phase is shown by the light-grey lines. Starting from 5.2~GPa, they coincide with the dark solid lines.}
	\label{f:structure-vs-pressure}
\end{figure*}

\subsection{B) Magnetic collapse}

The long-range magnetic order in our simulations is necessary to reproduce the main features of the experimentally discovered collapse transition around 4~GPa. However, the imposition of an artificial magnetic order in \CaK{}, which in reality has a superconducting ground state with no known, long range magnetic order, creates an additional feature at higher pressures which is solely due to the theoretically imposed long-range order and should not be (and is not) observed experimentally. Upon crossing the transition point at 4~GPa, the Fe magnetic moments become gradually suppressed and vanish completely at 5.2~GPa creating a further drop of the \textit{c} parameter (Fig.~\ref{f:structure-vs-pressure}\,(a)). For the magnetic and non-magnetic optimized structures, we calculated the pressure-dependent enthalpies with the same VASP parameters as described in the main text and analyzed the enthalpy difference between the two phases, in order to find a first-order transition point. This enthalpy analysis shows that the magnetic order is, in fact, destabilized already after 4.5~GPa, where the first-order transition between the twisted magnetic and the non-magnetic phase occurs. One can argue that the pressure range between 4.5~GPa and 5.2~GPa is the metastability region for the studied non-collinear magnetic configuration (indicated by the light-shaded region in Fig.~\ref{f:structure-vs-pressure}\,(a)).

To emphasize the role of the in-plane Fe spin orientation, we compare this twisted-order model to another configuration preserving the tetragonal symmetry, namely, the collinear checkerboard magnetic arrangement (Fig.~\ref{f:supercells}\,(a)) in terms of its evolution under pressure. Our simulations of the collinear state revealed no abrupt changes in the lattice parameters within the same pressure range and show only gradual decrease of Fe magnetic moments upon reaching 5~GPa. The checkerboard model is, therefore, not suitable for describing the structural transitions in the \CaK{} compound.

\subsection{C) Half-collapse transitions near Ca and K}

\begin{figure*}
	\begin{center}
		\includegraphics[width=140mm]{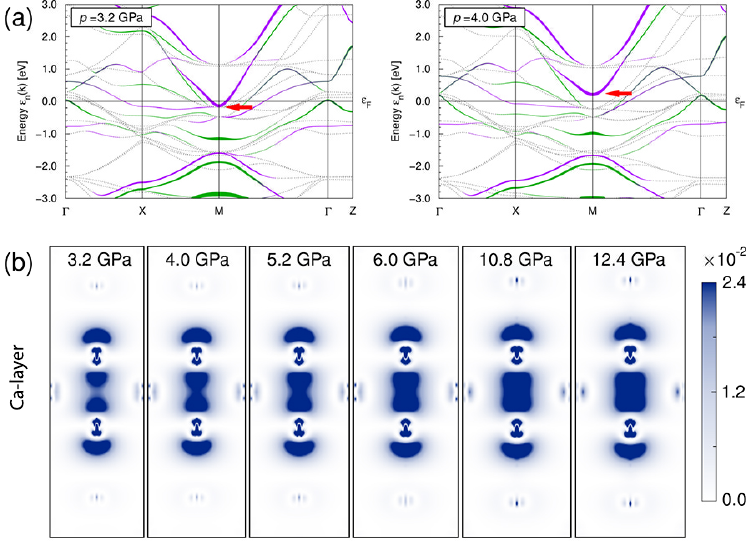}
		\includegraphics[width=140mm]{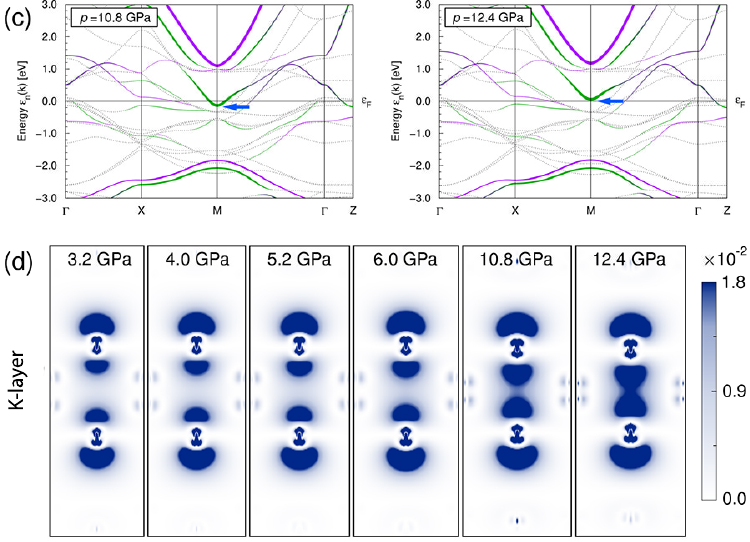}
	\end{center}
	\caption{(a,c) Evolution of the  non-spin polarized band structures with pressure across the collapse transitions on the Ca and K layers. Orbital weight of the As-$4p_z$ states near the Ca (violet) and K (green) layers is shown as well. The As-derived antibonding molecular orbital that experiences an upward shift upon crossing the transition point is marked by red and blue arrows for the Ca and K layers, respectively. (b,d) Non-spin-polarized electron density in the \textit{ac} plane associated to the As-$4p_z$ orbitals near Ca and K at different pressures. (b) shows clear bond formation across the Ca-layer by 4.0\,GPa and (d) shows clear bond formation across the K-layer by 12.4\,GPa. }
	\label{f:collapse_near_Ca_and_K}
\end{figure*}

For selected pressures, the non-spin polarized band structure  was calculated for the optimized lattice parameters and As positions using the all-electron full-potential localized orbitals basis set (FPLO) code~\cite{Koepernik1999PRB}. Analysis of the As-orbital weight near the Fermi energy allowed the identification of the respective collapse transitions. In particular, upon reaching the pressure of 4~GPa, the antibonding As-$4p_z$ molecular orbital near Ca is shifted above the Fermi level (Fig.~\ref{f:collapse_near_Ca_and_K}\,(a)) and, as a result, the corresponding As-As bonding becomes stronger. To illustrate this effect in the real space, we calculated the electron density maps associated to these orbitals by means of the tricubic interpolation~\cite{Lekien2005,pytricubic} of the Wannier functions obtained using the FPLO code\,\cite{Koepernik1999PRB}. On these maps, clear enhancement of the electron density between the two As ions is observed around 4~GPa (Fig.~\ref{f:collapse_near_Ca_and_K}\,(b)). As expected, further increase of pressure produces no qualitative changes in the orbital picture for this spatial region.

For As-As bonding across the K layer, on the other hand, the collapse occurs at much higher pressures, around 12~GPa. This fact is evidenced by the band structure plots (Fig.~\ref{f:collapse_near_Ca_and_K}\,(c)) where the As-derived molecular orbital near K crosses the Fermi level upon increasing pressure. Figure\,\ref{f:collapse_near_Ca_and_K}\,(d) makes it abundantly apparent that there is no qualitative change in As-As bonding across the K-layer anywhere near 4\,GPa.  Instead, the two As orbitals only clearly overlap across the K layer for $p$\,$\sim$\,12.4~GPa (Fig.~\ref{f:collapse_near_Ca_and_K}\,(d)). Thus, our computational results clearly indicate that the $\sim$\,4\,GPa transition is associated with a half-collapsed-tetragonal phase being stabilized.

\clearpage


\bibliographystyle{apsrev4-1}

%

	
\end{document}